\documentclass[journal=jctcce,manuscript=article]{achemso}
\usepackage{color}
\usepackage{amsmath}
\usepackage{graphicx}
\usepackage{ragged2e}
\usepackage{multirow}
\usepackage{multicol}
\usepackage{textcomp}
\usepackage{amssymb}
\usepackage[utf8]{inputenc}
\usepackage{gensymb}
\usepackage{lscape}
\usepackage{pdflscape}
\usepackage[table]{xcolor}
\usepackage{enumerate}   
\usepackage{titlesec}
\usepackage{braket}
\usepackage{graphicx}

\usepackage[T1]{fontenc}
\usepackage{amsfonts}

\usepackage{bm}

\usepackage{xspace} 
\usepackage{caption}
\usepackage{subcaption}
\usepackage{etoolbox}
\AtBeginEnvironment{tabular}{\footnotesize}
\usepackage{tikz}
\usetikzlibrary{arrows, arrows.meta}
\usepackage{comment}
\usepackage{newtxtext}
\usepackage{mathtools}
\setkeys{acs}{articletitle = true}

\newcommand*{\wfq}{$\omega$FQ\xspace}

\newcommand*{\wfqfmu}{$\omega$FQF$\mu$\xspace}

\newcommand*{\tqq}{\mathbf{T}^\mathrm{qq}}
\newcommand*{\tqmu}{\mathbf{T}^\mathrm{q\mu}}
\newcommand*{\tmuq}{\mathbf{T}^\mathrm{\mu q}}
\newcommand*{\tmumu}{\mathbf{T}^\mathrm{\mu\mu}}

\newcommand{\nextsubcolumn}[1][]{%
  \cr\noalign{\hfill}
  \if\relax\detokenize{#1}\relax\else\hsize=#1\setlength{\subcolumnwidth}{\hsize}\fi
}

\newlength{\subcolumnwidth}

\usepackage{xr-hyper}
\usepackage[hidelinks]{hyperref}

\usepackage[version=3]{mhchem} 

\author{Sveva Sodomaco}
\affiliation{Scuola Normale Superiore, Classe di Scienze, Piazza dei Cavalieri 7, 56126, Pisa, Italy}
\author{Piero Lafiosca}
\affiliation[SNS Pisa]{Scuola Normale Superiore, Classe di Scienze, Piazza dei Cavalieri 7, 56126, Pisa, Italy}
\author{Tommaso Giovannini}
\affiliation{Department of Physics and INFN, University of Rome Tor Vergata, Via della Ricerca Scientifica 1, 00133, Rome, Italy }
\alsoaffiliation{Consorzio Interuniversitario Nazionale per la Scienza e Tecnologia dei Materiali (INSTM), UdR Roma Tor Vergata, Via della Ricerca Scientifica 1, 00133, Rome, Italy}
\email{tommaso.giovannini@uniroma2.it}
\author{Chiara Cappelli}
\affiliation[SNS Pisa]{Scuola Normale Superiore, Classe di Scienze, Piazza dei Cavalieri 7, 56126, Pisa, Italy}
 \alsoaffiliation{Consorzio Interuniversitario Nazionale per la Scienza e Tecnologia dei Materiali (INSTM), UdR Pisa-SNS, Piazza dei Cavalieri 7, 56126 Pisa, Italy.}
\email{chiara.cappelli@sns.it}

\title[An \textsf{achemso} demo]
  {Atomistic QM/Classical Modeling of Surface-Enhanced Infrared Absorption      	
}

\begin{document}


\begin{abstract}
We present a multiscale quantum mechanics/classical (QM/MM) approach for modeling surface-enhanced infrared absorption (SEIRA) spectra of molecules adsorbed on plasmonic nanostructures. The molecular subsystem is described at the density functional theory (DFT) level, while the plasmonic material is represented using fully atomistic, frequency-dependent Fluctuating Charges (\wfq) and Fluctuating Charges and Dipoles (\wfqfmu) models. These schemes enable an accurate and computationally efficient description of the plasmonic response of both graphene-based materials and noble metal nanostructures, achieving accuracy comparable to \emph{ab initio} methods. The proposed methodology is applied to the calculation of SEIRA spectra of adenine adsorbed on gold nanoparticles and graphene sheets. The quality and robustness of the approach are assessed through comparison with surface-enhanced Raman scattering (SERS) spectra and available experimental data. The results demonstrate that the proposed framework provides a reliable route to simulate vibrational responses of plasmon–molecule hybrid systems.
\end{abstract}

\begin{tocentry}
\includegraphics[]{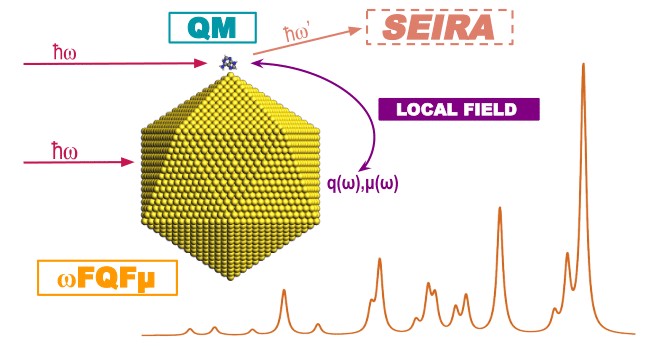}
\end{tocentry}

\begin{center}
\end{center}

\newpage

\section{Introduction}

Surface-enhanced vibrational spectroscopies, including surface-enhanced infrared absorption (SEIRA) and surface-enhanced Raman scattering (SERS), selectively amplify molecular vibrational signals by exploiting the intense local and inhomogeneous electric field enhancement near plasmonic nanostructures.\cite{lal2008tailoring, yang2018nanomaterial, hu2019graphene, minopoli2022nanostructured} SERS is among the most widely employed techniques in (bio)sensing due to its high sensitivity to the molecular fingerprints, which can be enhanced by several orders of magnitude, even allowing single molecule detection.\cite{langer2019present} SEIRA has found applications in diverse fields, including the in situ monitoring of metal surface catalytic reactions \cite{papasizza2018situ, heidary2019probing}, electrochemical studies \cite{alvarez2018situ, alvarez2019situ, alvarez2023guanine}, nanoparticle design \cite{quirk2016surface,piergies2021nanoparticle}, and analytical biosensing platforms \cite{yamada2004adsorption, kundu2009adenine,rodrigo2015mid,wagner2022towards}. 

First reported in 1980 via attenuated total reflection measurements,\cite{hartstein1980enhancement} SEIRA was initially studied for molecular monolayers adsorbed on metal films (e.g., Au, Ag, Cu, Pt), and on core–shell substrates.\cite{osawa1991surface,morton2011theoretical} These configurations typically yield enhancement factors up to $10^4$.\cite{hartstein1980enhancement,osawa1993surface}
More recently, the interest has shifted toward resonant SEIRA, which seeks to further improve sensitivity by engineering metal nanoantennas or graphene-based structures to make the plasmon resonance frequency (PRF) resonate with the vibrational modes of target molecules, i.e., in the infrared region.
\cite{neubrech2017surface} In particular, graphene and its derivatives have gained significant attention as SEIRA substrates due to their tunable plasmonic response in the mid-IR range.\cite{oh2021nanophotonic}
Similarly to SERS, the dominant enhancement mechanism in SEIRA arises from the local electromagnetic (EM) mechanism at the molecule–substrate interface, which is primarily linked to the excitation of the localized surface plasmon. However, while SERS enhancement depends on the fourth power of the local field, SEIRA exhibits a quadratic dependence, yielding intrinsically lower intensities as compared to SERS.\cite{lal2008tailoring} 

Theoretical modeling plays a crucial role in predicting and interpreting SEIRA and SERS spectra.\cite{morton2011theoretical} Full Quantum Mechanical (QM) approaches have typically considered molecules adsorbed on small noble metal clusters.\cite{kryachko2005complexes,kundu2009adenine,pan2022atomically}
However, such models fall short in reproducing the optical response properties of realistic plasmonic substrates.\cite{morton2011theoretical,zhang2018size} To overcome this limitation, multiscale approaches have been developed that couple a QM description of the molecule with classical electromagnetic models of the plasmonic environment. Continuum models, such as the Boundary Element Method (BEM),\cite{de2002retarded} which represent the nanostructures as a continuum dielectric, have been adopted for this purpose.
\cite{corni2001enhanced,corni2001theoretical,corni2002surface,Corni2006,morton2011theoretical,takenaka2020combined,illobre2025mixed} Nevertheless, to fully account for atomistic details such as edges, tips, defects, and junctions, which are critical to achieving high local field enhancement, atomistic methodologies must be exploited.\cite{morton2011discrete,payton2014hybrid,urbieta2018atomic,zakomirnyi2019extended,bonatti2020plasmonic,pei2022analytic,lafiosca2023qm,giri2024modeling} In this framework, we have recently introduced a multiscale QM/classical approach for simulating SERS spectra,\cite{lafiosca2023qm} based on the fully atomistic frequency-dependent fluctuating charges ($\omega$FQ)\cite{giovannini2019classical,giovannini2020graphene} and dipoles ($\omega$FQF$\mu$)\cite{giovannini2022we} approaches. These models, rooted in the Drude conduction theory with phenomenological corrections for quantum tunneling, enable the simulation of the optical response of a wide range of plasmonic materials, including alkali metals,\cite{giovannini2019classical} noble metal nanoparticles,\cite{giovannini2022we, nicoli2023fully, nicoli2024atomistic} nanoalloys,\cite{nicoli2023fully} and graphene-based substrates,\cite{giovannini2020graphene} even in the presence of structural defects,\cite{bonatti2022silico,zanotto2023strain} solvent effects,\cite{nicoli2024atomistic} and subnanometer gaps.\cite{giovannini2019classical,giovannini2022we,giovannini2025electric} Their classical formulation allows simulations of systems with more than one million atoms,\cite{lafiosca2021going} yielding an almost perfect agreement with full \textit{ab initio} methods,\cite{giovannini2019classical,bonatti2020plasmonic,giovannini2022we} while retaining a fully atomistic resolution.

In this work, we extend the QM/$\omega$FQ(F$\mu$) approach to SEIRA spectroscopy and apply it to adenine, one of the DNA nucleobases, adsorbed on gold nanostructures and graphene disks. Adenine represents a particularly relevant test system for SEIRA studies, as it has been extensively investigated both experimentally and theoretically.\cite{kundu2009adenine,rueda2012situ,acres2018experimental,harroun2018controversial,hu2019surface} SEIRA spectra of adenine have been reported on a variety of gold-based substrates, including Au(111) and Au(100) electrodes,\cite{rodes2009adenine,rueda2012situ,alvarez2014situ,alvarez2018situ} as well as silica core–gold nanoshells.\cite{kundu2008surface} More recently, SEIRA on graphene oxide (GOEIRA) has also been explored, highlighting the potential of carbon-based platforms.\cite{hu2019selective} 

The paper is organized as follows. First, the computational framework used to investigate SEIRA and SERS spectra is introduced. The approach is then applied to predict the SEIRA signals of adenine (ADE) nucleobase in proximity to gold nanostructures, with a focus on the influence of molecular orientation and adsorption site on the spectral response. The results are compared with SERS spectra and validated against available experimental data. In the final section, we assess the potential of graphene as SEIRA substrate by analyzing how its structural and electronic properties modulate the molecule-graphene response in the IR range. Summary, conclusions, and future perspectives end the paper.

\section{Theory}

SEIRA spectra are computed using a multiscale QM/classical approach. In this framework, the plasmonic nanostructure and its interaction with external radiation, leading to the formation of localized surface plasmons, are described through the fully atomistic \wfq{} and \wfqfmu{} models, whereas the molecular subsystem is treated quantum mechanically at the DFT and TDDFT levels. In this section, we briefly recall the theoretical foundations of \wfq{} and \wfqfmu{}, and then present their coupling to a QM region and its extension to SEIRA simulations.

\subsection{\texorpdfstring{$\omega$FQ and $\omega$FQF$\mu$}{omegaFQ and omega FQFmu} models for nanoplasmonics}

$\omega$FQ is a classical, fully atomistic, and frequency-dependent model able to describe the plasmonic features of various materials, including alkali metals \cite{giovannini2019classical}, and graphene-based materials \cite{giovannini2020graphene}. In $\omega$FQ, each atom is endowed with a complex electric charge, and the charge exchange between the atoms is assumed to be governed by a Drude-conduction mechanism mediated by a phenomenological damping, which limits the charge flow between nearest neighbors and mimics the physics of quantum tunneling.\cite{giovannini2019classical} The charges are obtained by solving the following linear system defined in the frequency domain:\cite{lafiosca2021going}
\begin{equation}\label{eq:wfq}
\left[\overline{\mathbf{K}}\tqq-z_q(\omega)\mathbf{I}_N\right]\mathbf{q}(\omega) = -\overline{\mathbf{K}}\mathbf{V}^{\text{ext}}(\omega)
\end{equation}
where $\overline{\mathbf{K}}$ is a $N \times N$ matrix whose elements read:
\begin{equation}
    \label{eq:kappa_bar}
    \overline{K}_{ij} = K_{ij} - \sum_k K_{ik}\delta_{ij}
\end{equation}
where $\delta_{ij}$ is the Kronecker delta. The elements of the $\mathbf{K}$ matrix take the following form:
\begin{equation}\label{eq:drude_matrix}
K_{ij} = 
\begin{cases}
\frac{[1-f(r_{ij})]\mathcal{A}_i}{r_{ij}} & \mbox{if } i\ne j \\ 
0 & \mbox{if } i = j \\
\end{cases}
\end{equation}

The left-hand side of Eq.\ref{eq:wfq} is composed of a real frequency-independent matrix with a diagonal shift of the complex scalar frequency-dependent $z_q(\omega)$, which is a function of the relaxation time $\tau$, a friction-like constant that accounts for scattering events, and the electron density of the system, $n$. $\mathbf{I}_N$ is the identity matrix of order $N$, where $N$ is the number of atoms.
The charge flow between atoms results from the Drude conduction mechanism, the interaction between charges, represented by the $\tqq$ interaction kernel \cite{giovannini2019classical}, and with the external potential $\mathbf{V}^\mathrm{ext}$ associated with the external electric field oscillating at frequency $\omega$.
$K_{ij}$ is a real symmetric Drude matrix, expressed in terms of the effective area $A_{i}$ of atom $i$. $f(l_{ij})$ is a Fermi-like function effectively modeling quantum tunneling effects ($l_{ij}$ is the distance between atoms $i$ and $j$).

$\omega$FQ can be extended to treat graphene-based nanomaterials,\cite{giovannini2020graphene} by introducing the effective mass $m^*$ (which is 1 a.u. for pure metals)\cite{neto2009electronic}. The 3D atomic effective electron density of graphene $n_{0}$ reads:
\begin{equation}
     n = \frac{\tilde{n}_{0}}{m^{*}} = \sqrt{\frac{n_{2D}}{\pi}}a_{0}v_{F} = \frac{E_{F}a_{0}}{\hbar \pi} \: 
     \label{eq:dens_graph}
\end{equation} 
where $\tilde{n}_{0}$ is the the 3D atomic electron density, $v_{F}$ is the Fermi velocity, $E_{F}$ is the Fermi energy, $a_{0}$ is the Bohr radius, and $n_{2D}$ is the two-dimensional electron density of graphene.\cite{neto2009electronic,giovannini2020graphene} In Eq. \ref{eq:dens_graph}, we have used the relationship between $E_{F}$ and $n_{2D}$: $E_{F}=\hbar v_{F} \sqrt{\pi n_{2D}}$.\cite{neto2009electronic,giovannini2020graphene} Eq. \ref{eq:dens_graph} shows that, for graphene-based nanostructures, the Drude dynamics also depends on the Fermi energy, which can be tuned experimentally \cite{koppens2011graphene,xu2011effect,garcia2014graphene,rodrigo2015mid,valevs2017enhanced,giovannini2020graphene, bonatti2022silico,zanotto2023strain}.\\

While \wfq{} can properly describe alkali metals and graphene, it fails to account for the optical response of noble metals, as it assumes that the conduction electrons follow a purely Drude-like behavior and neglects interband transitions.\cite{pinchuk2004influence,pinchuk2004optical,balamurugan2005evidence} To overcome this limitation, we introduce an additional polarization source that accounts for the polarizability of the $d$-shell.\cite{liebsch1993surface} In the resulting \wfqfmu approach,\cite{giovannini2022we} each atom is represented by a charge $q$ and a dipole $\bm{\mu}$, the latter depending on the interband frequency-dependent atomic polarizability $\alpha_{IB}(\omega)$. The equations of motion of charges and dipoles are coupled by solving the following linear system of equations:

\begin{equation}
\label{eq:wfqfmu_eq_nocompact}
\resizebox{\textwidth}{!}{$
\left[
\begin{pmatrix}
\overline{\mathbf{K}} & \mathbf{0} \\ 
\mathbf{0} & \mathbf{I}_{3N}
\end{pmatrix}
\begin{pmatrix}
\tqq & \tqmu \\ \tmuq & \tmumu
\end{pmatrix}
-\begin{pmatrix}
z_q(\omega)\mathbf{I}_N & \mathbf{0} \\ \mathbf{0} & z_\mu(\omega)\mathbf{I}_{3N}
\end{pmatrix}
\right]
\begin{pmatrix}
\mathbf{q}(\omega) \\ \bm{\mu}(\omega)
\end{pmatrix}
= 
\begin{pmatrix}
\overline{\mathbf{K}} & \mathbf{0} \\ 
\mathbf{0} & \mathbf{I}_{3N}
\end{pmatrix}
\begin{pmatrix}
-\mathbf{V}^{\mathrm{ext}}(\omega) \\ \mathbf{E}^{\mathrm{ext}}(\omega)
\end{pmatrix}
$}
\end{equation}
where 
\begin{equation}
    z_{\mu}(\omega) = - \frac{1}{\alpha_{IB}(\omega)} \: 
\end{equation}
The left-hand side of the linear system in Eq. \ref{eq:wfqfmu_eq_nocompact} comprises all the interaction kernels: charge-charge $\tqq$, charge-dipole $\tqmu$, dipole-charge $\tmuq$, and dipole-dipole $\tmumu$. The right-hand side of the system contains the external sources, i.e., the potential ($\mathbf{V}^\mathrm{ext}$) and the electric field ($\mathbf{E}^\mathrm{ext}$).

\subsection{QM/\texorpdfstring{$\omega$FQF$\mu$}{omega FQ(Fmu)} for SEIRA}

QM/\wfqfmu has recently been extended to the damped linear response formalism to compute complex polarizabilities, from which SERS signals can be calculated.\cite{lafiosca2023qm} All the technical details on the derivation of the formalism in the time-dependent KS (TDKS) framework, as well as the details of the ground-state (GS) coupling, can be found in Ref. \citenum{lafiosca2023qm}. 

The QM/\wfqfmu{} approach for SERS spectra assumes that the perturbation operator includes both the electric potential of the external field and the local field operator accounting for the plasmon-induced field generated by the substrate.\cite{corni2001theoretical,payton2014hybrid} These effects are incorporated into the effective KS operator through the image field term, describing the polarization induced by the nanostructure’s charges and dipoles, and the local field term due to the plasmon, both evaluated self-consistently with the perturbed TDKS density.\cite{corni2001theoretical,payton2014hybrid} Solving the coupled-perturbed TDKS equations yields the frequency-dependent complex molecular polarizability tensor, which naturally includes the scattered and reflected fields responsible for the electromagnetic enhancement observed in SERS. A similar physical picture is followed here to extend QM/\wfqfmu to SEIRA intensities. In particular, in line with Ref.\citenum{cammi2000calculation}, IR absorption intensities can be calculated from the geometrical derivatives of an "external" dipole moment, which is the sum of the gas-phase molecular dipole moment, $\bm{d}$, and an additional dipole moment, $\tilde{\bm{d}}$, induced on the nanostructure by the molecular density, i.e. that acccounts for local field effects.\cite{cammi1998calculation,tomasi1999medium} The induced dipole $\tilde{\bm{d}}$ can be expressed as:
\begin{equation}\label{eq:dipole}
-\tilde{\bm{d}}\cdot\mathbf{E}^\mathrm{ext} = \sum_{p=1}^{N} \biggl[ q_p V^\mathrm{ext}(\mathbf{r}_p) - \bm{\mu}_p\cdot\mathbf{E}^\mathrm{ext}(\mathbf{r}_p) \biggr]
\end{equation}
where $q_p$ and $\bm{\mu}_p$, located at position $\mathbf{r}_p$, represent charges and dipoles induced by the GS QM density on the plasmonic substrate.
%
These are calculated through Eq.\ref{eq:wfqfmu_eq_nocompact} by substituting the external field with the electric potential $\mathbf{V}^{QM}$ and field $\mathbf{E}^{QM}$ generated by the oscillating molecular density. 
Thus, the induced dipole is obtained from the GS QM electric potential and field, and the local field charges and dipoles oscillating at the normal mode frequency.

\section{Computational details}

QM/$\omega$FQ(F$\mu$) s applied to the calculation of SERS and SEIRA spectra of the adenine (ADE) nucleobase adsorbed on gold and graphene substrates. ADE is chosen because it is of broad interest in biosensing due to its role as a DNA building block.\cite{ataka2007biochemical,barhoumi2008surface,brown2008nucleotide,kundu2009adenine,liu2012adsorption} The gold nanoparticle is characterized by an icosahedral morphology, comprising 10179 atoms (Au$_{10179}$ Ih) with a radius of about 3.83 nm and 14 gold shells, and its dipolar plasmon resonance frequency (PRF) is 2.21 eV (calculated at the \wfqfmu level). Two additional gold substrates, Au$_{49049}$ Ih and Au$_{104223}$ Ih, are considered, with PRFs falling at 2.21 eV and 2.18 eV, respectively, to showcase the dependence of SERS and SEIRA spectra by enlarging the nanostructure. For graphene substrates, we consider graphene disks, with geometries taken from Refs. \citenum{bonatti2022silico} and \citenum{lafiosca2023qm}, characterized by diameters ranging from 24 nm (17269 atoms) to 100 nm (300695 atoms) and a Fermi energy of 0.40 eV. In addition, we examine a 32 nm graphene disk for which the Fermi energy is tuned from 0.40 to 0.09 eV. 
The gold and graphene substrates are treated at the \wfqfmu and \wfq levels by exploiting the parameters from Refs. \citenum{giovannini2022we} and \citenum{giovannini2020graphene}, respectively. 

\begin{figure}[!htbp]
\centering
    \centering
    \includegraphics[width=0.3\textwidth]{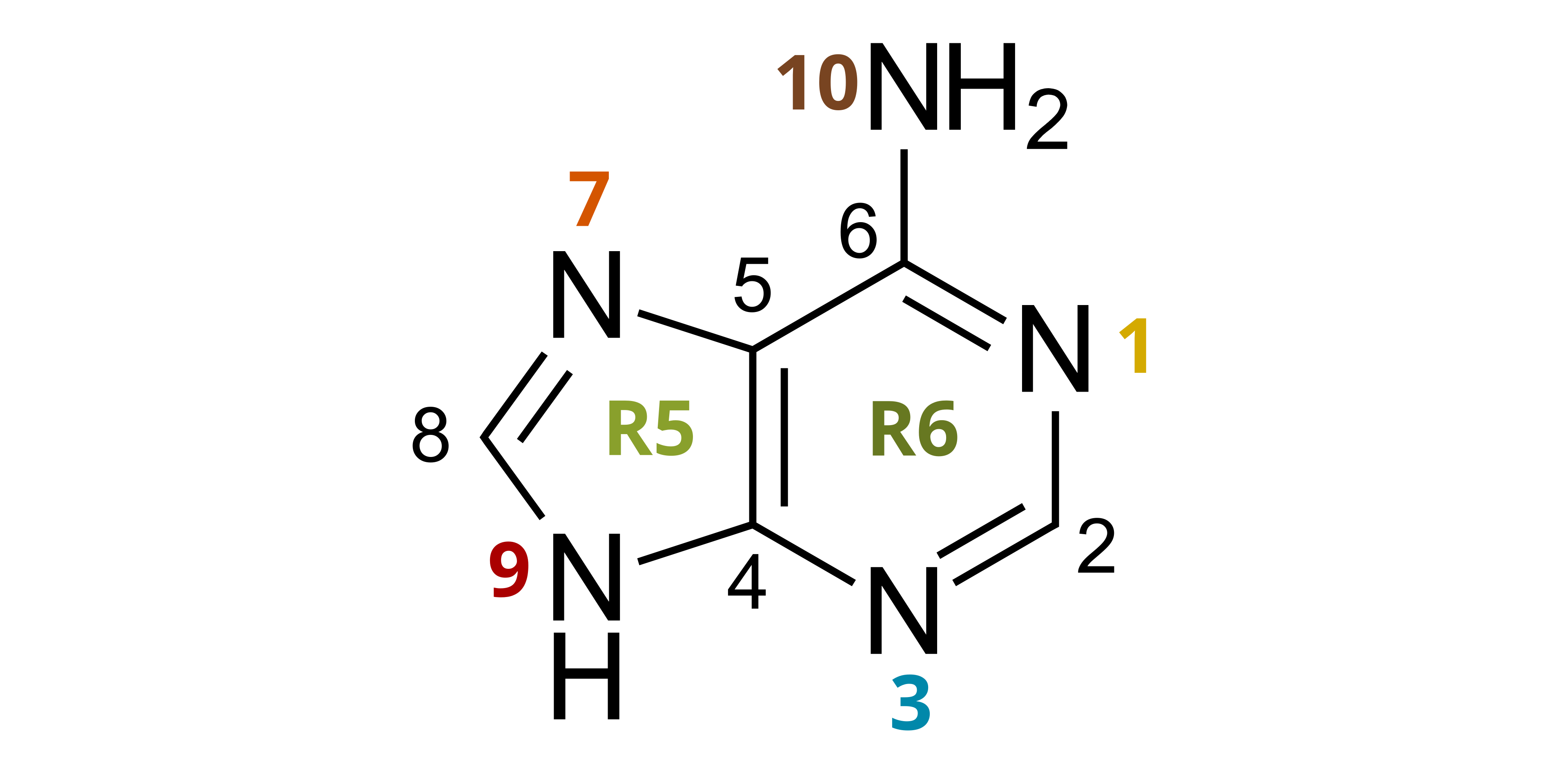}
    \caption{Molecular structure and atom labelling of adenine.}
    \label{fig:ade_label}
\end{figure}

In all molecule-substrate configurations, the 9H tautomer of adenine (ADE), shown in Fig.\ref{fig:ade_label}, is described at the DFT level by using the BP86 functional and the TZP basis set, in agreement with previous studies\cite{payton2014hybrid,lafiosca2023qm}, with the default DFT integration grid for adenine on gold and a finer grid for adenine on graphene (see also Fig. S1a in Supporting Information -- SI).
The distance between ADE Nitrogen atom and the closest gold atom of the NP is set to 3 \AA~. 
For specific configurations, v\_N9, v\_N10, f\_N1, f\_N7, and f\_N10, the N–Au distance is increased to 3.4 \AA~to avoid steric interactions between molecular Hydrogen and Au atoms.

Normal mode displacements are obtained from the isolated QM molecule since the effect of the plasmonic substrate is negligible.\cite{payton2014hybrid,lafiosca2023qm} Geometrical derivatives of frequency-dependent dipoles (SEIRA) and polarizabilities (SERS) are obtained using a two-point numerical procedure with a differentiation step of 0.001 Bohr. For completeness, Fig. S1b in the SI reports a numerical stability test carried out with a step size of 0.0005 Bohr. 
SERS spectra are simulated by setting the Raman incident field at the PRF of the substrate, and we use a lifetime of 0.10 eV, in agreement with previous studies.\cite{payton2014hybrid,lafiosca2023qm} 
SERS and SEIRA raw data are convoluted by using a Lorentzian band shape with full width at half maximum (FWHM) of 10 cm$^{-1}$. All QM/$\omega$FQ and QM/$\omega$FQF$\mu$ calculations are performed by using a locally modified version of the AMS software \cite{ams2020, baerends2025amsterdam}.
By following the procedure reported in Ref.\citenum{lafiosca2023qm} for SERS, QM/FQ(F$\mu$) contributions to the GS were neglected, as their effect on SEIRA signals is only marginal (see also Fig. S2 in the SI).

To quantify spectral enhancements due to the presence of the plasmonic nanostructure,\cite{lafiosca2023qm}, we introduce the averaged enhancement factor (AEF) and the maximum enhancement factor (MEF), which are computed from the IR/Raman enhancement factor (EF) associated with each $i$-th normal mode: 
\begin{equation}
     EF^i(\omega) =  \frac{I^i(\omega)}{I_{vac}^i(\omega)}; \quad AEF(\omega)=\frac{\sum_{i}I^i(\omega)}{\sum_{l}I_{vac}^l(\omega)} ; \quad MEF(\omega)=\max\limits_i EF^i(\omega) \: .
\end{equation}
The $i$-th normal mode showing maximum enhancement factor is indicated as $i$-MEF.

\section{Numerical Results} 

\subsection{SEIRA and SERS spectra of adenine on gold nanostructures}

\begin{figure}[!htbp]
    \centering
    \includegraphics[width=0.5\textwidth]{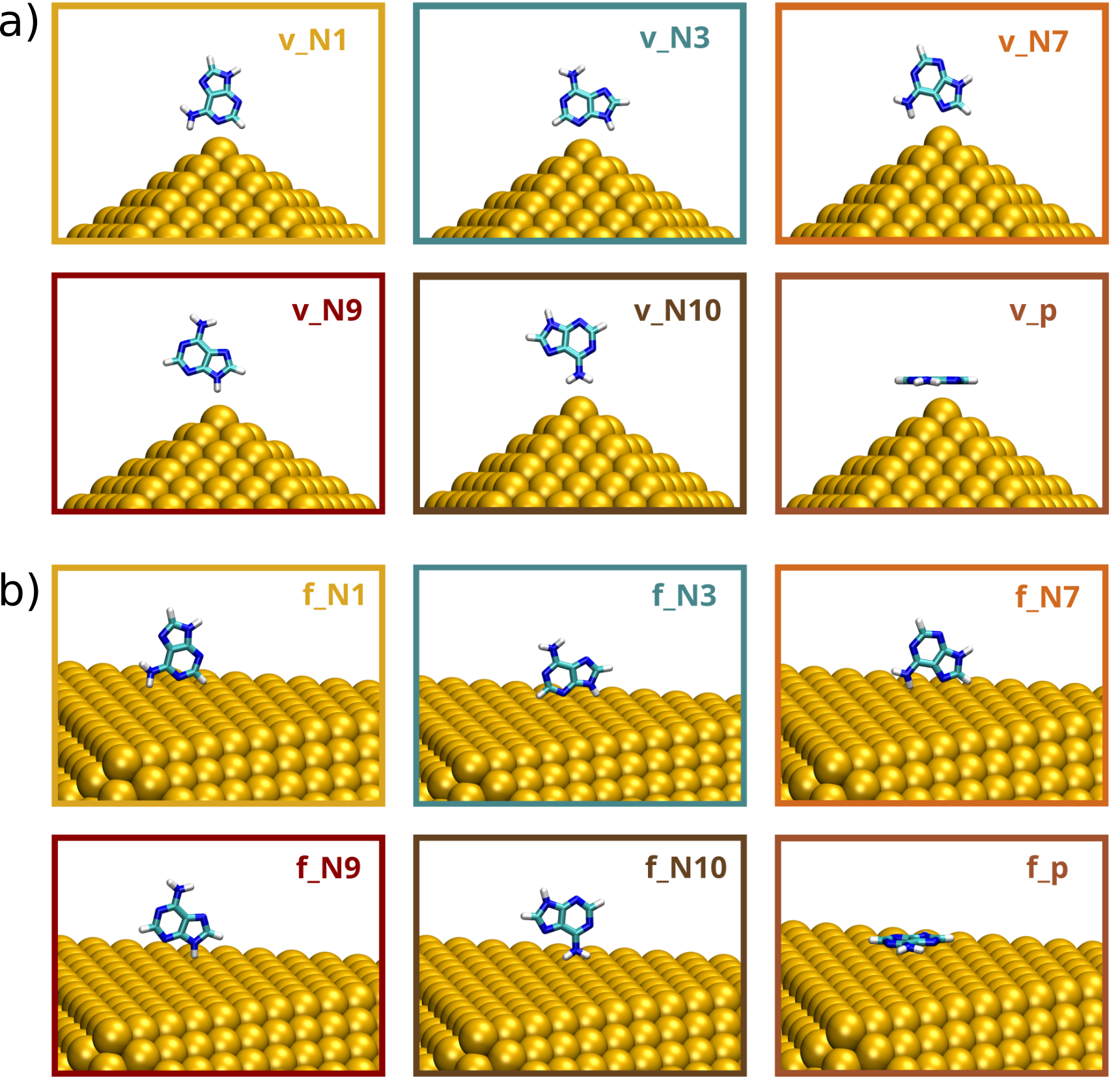}
    \caption{Molecular configurations of ADE adsorbed on the vertex (a) and face (b) positions of Au$_{10179}$ Ih NP.}
    \label{fig:geometries_ade_gold}
\end{figure}

\begin{figure}[H]
    \includegraphics[width=1\textwidth]{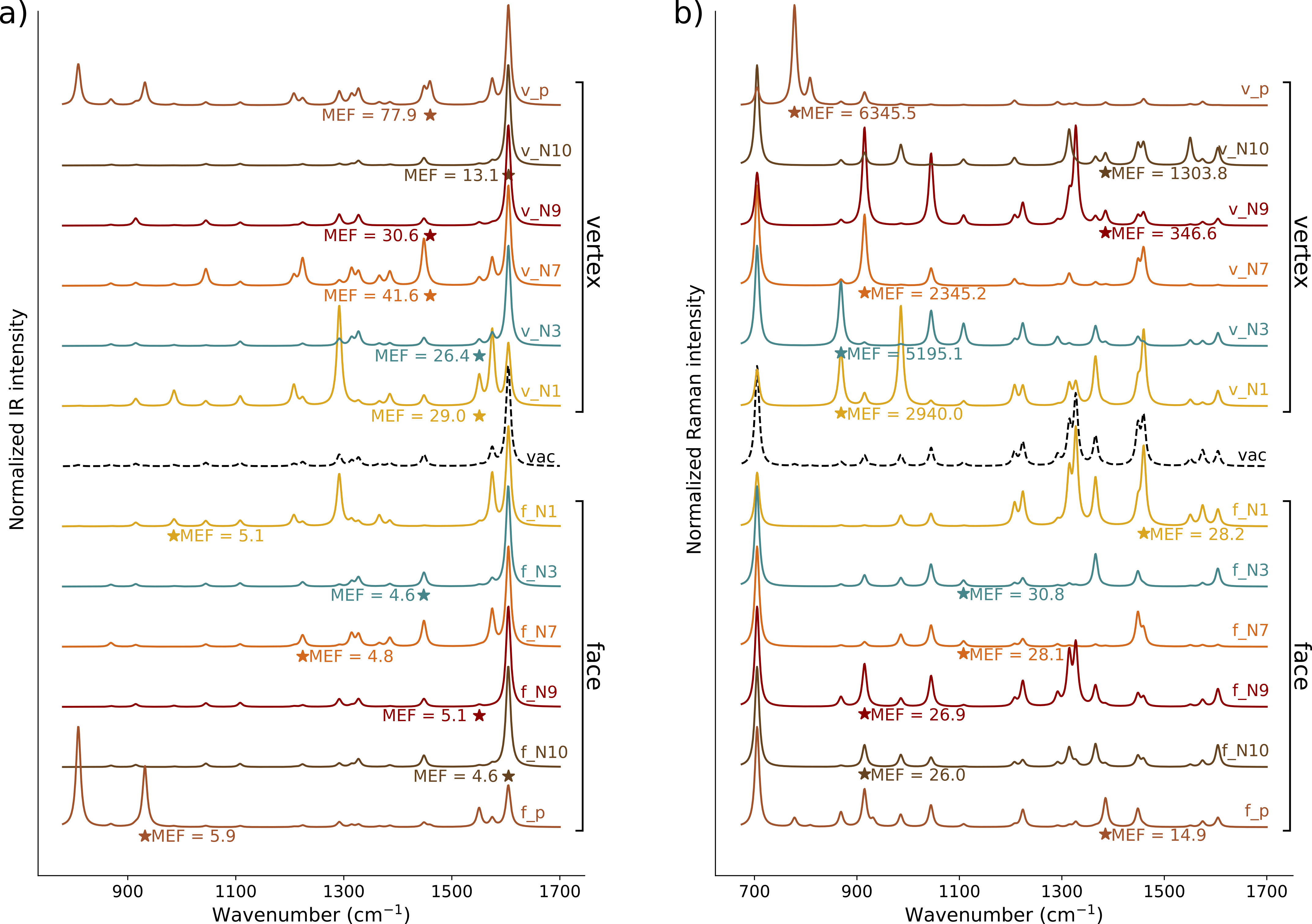}
    \caption{Normalized SEIRA (a)) and SERS (b)) spectra of the six configurations of ADE adsorbed on the vertex and face of Au$_{10179}$ Ih. SERS spectra are calculated at the PRF of the gold substrate (2.21 eV/560 nm). Stars denote $i$-MEF. MEF values are also reported.}
\label{fig:seira_sers_ade}
\end{figure}

Multidentate binding modes render the preferred orientation of ADE on gold and silver surfaces challenging to resolve.\cite{harroun2018controversial} In fact, in its 9H tautomeric form, adenine has five nitrogen atoms as potential coordinating sites for metal surfaces. Several adsorption models for adenine have been proposed in the literature.\cite{harroun2018controversial} Adsorption \textit{via} the N3/N9 side (see Fig. \ref{fig:ade_label} for atom labelling) is supported by combined SERS and SEIRA spectra on gold nanoshells \cite{kundu2009adenine}, electrochemical tip-enhanced Raman spectroscopy (EC-TERS) studies on protonated adenine \cite{martin2017electrochemical}, X-ray absorption and DFT calculations \cite{acres2018experimental}, and SERS spectra on gold nanoparticles \cite{yoshimoto2022three}. Adsorption \textit{via} N7/N10 sites is backed by SEIRA and cyclovoltammetry studies on gold electrodes \cite{rodes2009adenine, rueda2012situ, alvarez2014situ}, where surface-enhanced ring stretching modes (N7-C5, N7-C8) and scissoring of the amino group were observed. Similarly, the N1/N10 adsorption model is supported by subtractively normalized interfacial Fourier transform infrared spectra (SNIFTIRS) on Au(111) \cite{prieto2016quantitative} and EC-TERS studies for deprotonated adenine at higher potentials, suggesting a tilted flat orientation \cite{martin2017electrochemical}.
Although recent SERS and SEIRA experimental evidence proposes a vertical or tilted orientation of adenine on gold \cite{yoshimoto2022three}, computational studies, relying on molecular dynamics simulations, indicate that a flat orientation may occur at low concentrations.\cite{rosa2014enthalpy, sodomaco2023computational} To account for all possible adsorption sites, in this work we consider six binding modes of ADE on gold: four ``end-on'' modes involving each purine ring nitrogen atoms (N1, N3, N7, and the protonated N9), one ``face-on'' mode with the adenine rings oriented parallel to the surface (p), and one mode involving the exocyclic amino group (N10).\cite{kundu2009adenine} To investigate the influence of these binding orientations at different adsorption sites on a model Au$_{10179}$ Ih nanostructure, 12 configurations are generated, six at the vertex and six at the face of the gold nanostructure, as shown in Fig.\ref{fig:geometries_ade_gold}a) and b). The corresponding structures are named using the following nomenclature: $v\_{atom}$ (vertex) or $f\_{atom}$ (face), where $atom$ indicates ADE atom binding to the metal surface (see Fig. \ref{fig:ade_label} for atom labelling). We then exploit QM/\wfqfmu to analyze the sensitivity of SEIRA and SERS signals to the specific adsorption sites of various ADE orientations on the Au NP. The resulting spectra are graphically reported in Fig.\ref{fig:seira_sers_ade}, where they are compared to IR and Raman spectra of the isolated molecule, which are used as a reference to quantify the surface-induced variations.

In the gas phase, the ADE IR spectrum (black dashed line in Fig.\ref{fig:seira_sers_ade}a)) features four main peaks: the most intense (1605 cm$^{-1}$) corresponds to NH$_{2}$ scissoring and C5-C6 and C6-N10 stretching modes, and is accompanied by a shoulder peak at 1575 cm$^{-1}$, which is associated with N9-H and C8-H bending and N3-C4, N1-C6, C5-N7 and N7-C8 stretching (see Fig. S3 in the SI for a graphical depiction of ADE normal modes). Less intense bands are present at 1448 cm$^{-1}$, assigned to the stretching of N7-C8, N1-C6, and C2-N3 and the bending of C2-H and C8-H, and 1292 cm$^{-1}$, corresponding to the stretching of C2-N3 and C5-N7, together with the bending of C2-H, C8-H, and N9-H. 

Most SEIRA spectra closely resemble the gas-phase IR spectrum, especially v\_N3, v\_N9, v\_N10 (vertex configurations), and f\_N3, f\_N9, f\_N10 (face configurations), suggesting that the IR-active vibrational modes of ADE are largely preserved upon adsorption. All configurations exhibit their most intense peak at 1605 cm$^{-1}$, except for v\_N1 and f\_p configurations. In particular, v\_N1 presents the dominant peak at 1292 cm$^{-1}$, which corresponds to an in-plane normal mode involving stretching of the C2–N3 and C5–N7 bonds, along with bending motions of the C2–H, C8–H, and N9–H bonds. In contrast, for f\_p, the strongest signals correspond to out-of-plane modes at 808 cm$^{-1}$ (C8–H wagging) and 932 cm$^{-1}$ (C2–H wagging), the latter also reporting the MEF (5.9, see Tab. S1 in the SI). Such bands are also particularly intense for v\_p configuration. The f\_N1 configuration also deviates from the typical pattern, displaying two intense peaks at 1292 cm$^{-1}$ and 1575 cm$^{-1}$, linked to pronounced deformations of the adenine rings. Remarkably, the spectra obtained for ADE adsorbed on the vertex or face binding the same atom to the metal surface display a similar spectral profile, suggesting a relevant role of the binding site. To further deepen this point, in Fig. S4 in the SI, we report the SEIRA spectrum as obtained by simply averaging the spectra over the 6 configurations on the vertex and the face of Au$_{10179}$ Ih. The two obtained spectra are characterized by the same profile, confirming that the adsorption site (face or vertex) does not significantly alter the overall spectral shape. 

The face and vertex adsorption configurations are, however, associated with substantial differences in the enhancement factors. In fact, as can be appreciated by Fig. 3a and Tab. S1 in the SI, the MEFs calculated for ADE adsorbed on the vertex of the gold icosahedral are generally one order of magnitude larger than the corresponding values computed for face dispositions. This is not surprising and is related to the so-called tip effect, which arises from the highly localized and inhomogeneous electric fields at metallic nanostructures ``hot spots'' (sharp tips and vertices).\cite{jackson1999electrodynamics,lal2008tailoring} The normal modes associated with MEF ($i$-MEF) in most configurations are in-plane vibrations (see Tab. S1 in the SI), which is expected considering that the NP induced electric field is supposed to display the largest variations perpendicularly to the gold surface. It is also worth noting that the most intense band of the spectrum is generally not associated with the normal mode reporting the largest MEF, except for v\_N10 (MEF = 13.1) and f\_N10 (MEF = 4.6).

An enhancement of the IR signal (AEF) is observed across all configurations except for v\_p, where the IR signal is overall quenched (AEF < 1, see Tab. S1 in the SI). For the latter structure, a strong enhancement is detected at 1460 cm$^{-1}$ (involving the NH$_2$ group) which also dominates the MEF profile in v\_N7 and v\_N9 (see Fig. S5 in the SI). This suggests a particularly favorable orientation of the NH$_2$ oscillating dipole relative to the local electromagnetic field. Notably, the MEF of the 1460 cm$^{-1}$ band in these three configurations is the largest across all considered geometries. The raw data reported in Tab. S1 in the SI also highlights that SEIRA spectra for face configurations show lower AEFs compared to vertex configurations, coherently with the comment above regarding the MEFs. Interestingly, by plotting the enhancement factors for each normal mode (see Fig. S5b in the SI), we note that, for the face configurations, the enhancement factors are uniform across vibrational modes, clustering around the AEF, with the $i$-MEFs depending on the specific binding pose of the molecule. On the contrary, the vertex configurations plots (Fig. S5a in the SI) are dominated by the specific normal modes that report the MEF. This is again a consequence of the so-called ``tip effect''.
Overall, our calculated AEFs average around 5, with v\_N10 showing the highest value (9.2), and MEFs reach approximately 30, which aligns well with the typical SEIRA enhancement range, generally expected on the order of 10–100.\cite{hartstein1980enhancement,osawa1991surface,osawa1993surface}


We now move on to comment on the SERS signals of ADE adsorbed on Au NP. The gas phase Raman spectrum (see Fig. \ref{fig:seira_sers_ade}b) is characterized by a dominant peak at 705 cm$^{-1}$, corresponding to the ``breathing'' of the purine rings, along with characteristic peaks in the 1300–1500 cm$^{-1}$ region,\cite{mathlouthi1984ft} primarly in-plane bending normal modes, most of which involve the NH$_2$ group. In Fig. \ref{fig:seira_sers_ade}b), the SERS spectra of each adsorption configuration are graphically reported. Each vertex geometry displays a distinct most intense peak: 986 cm$^{-1}$ for v\_N1, 705 cm$^{-1}$ for v\_N3, v\_N7, and v\_N10, 1327 cm$^{-1}$ for v\_N9, and 778 cm$^{-1}$ for v\_p. Differently, most face configurations report the most intense peak at 705 cm${^-1}$ (ring breathing), with the exception of f\_N1. The f\_N1 SERS spectrum is in fact dominated by peaks in the region 1200-1500 cm$^{-1}$, with the most intense peak appearing at 1327 cm$^{-1}$ (in-plane stretching mode involving C8–N9, C6–N1, and N3–C4 bonds, along with bending of C8–H and N9–H). This highlights a significant difference with respect to SEIRA (see Fig. \ref{fig:seira_sers_ade}a). In fact, in SEIRA, the spectral profile is mainly determined by the binding atom. The same is not valid for SERS, for which face and vertex spectra of configurations exploiting the same binding site substantially differ. Furthermore, the SERS spectra of face structures mostly resemble the spectral profile of ADE in vacuo. Figure S6b in the SI further illustrates this trend by plotting the enhancement factors as a function of the normal mode. For face configurations, these are relatively uniform across Raman bands, reflecting a similar behavior as in SEIRA. On the contrary, the SERS spectra of vertex structures substantially deviate from the gas-phase, also activating quasi-inactive normal modes in the gas phase. As an example, we highlight the peak at 778 cm$^{-1}$ (out-of-plane ring deformation) for the v\_p configuration, for which the highest MEF (6345) is reported. This can also be appreciated in Fig. S6b in the SI, where the vertex configurations display a selective enhancement profile as a function of the normal mode, with EF values varying by up to two orders of magnitude. To further highlight the effects of the adsorption site, as for SEIRA, the SERS spectra averaged on the 6 face and 6 vertex configurations, separately, are reported in Figure S4 in the SI. Different from SEIRA, the two averaged spectra show significant discrepancies, highlighting, in this case, the substantial influence of the nanostructure-molecule morphology and the adsorption site on the spectral profile. 

The differences between face and vertex configurations are also reflected in the calculated values of AEF and MEF. As for SEIRA, the SERS enhancement factors are generally lower when ADE is adsorbed on the face than on the vertex. In fact, AEF and MEF values for vertex configurations are respectively one and two orders of magnitude higher than those observed for face configurations. In most configurations, the MEF is associated with normal modes involving in-plane atom displacements, as expected considering the electric field gradient induced by the plasmon excitation in the nanostructure. As mentioned above, the only exception is v\_p, which is characterized by a flat adsorption on the Au NP vertex, thus enhancing mostly out-of-plane normal modes, although providing an overall small AEF (47.6). The average AEF and MEF values are one and two orders of magnitude higher than those observed for SEIRA, and are of the order of 10$^2$ and 10$^3$ (vertex) and 10$^1$ and 10$^2$ (face). This is consistent with the electromagnetic enhancement theory, for which in SERS, the local electromagnetic enhancement scales with the fourth power of the local induced field, whereas in SEIRA, it scales with the square. 


We finally note that the absolute values of AEF and MEF of vibrational spectroscopies depend on the nanoparticle size, because of the larger induced field in the proximity of the nanostructure surface.\cite{kelly2003optical, willets2007localized,giovannini2022we} To showcase such a dependence, in Fig. S8 in the SI, we report the SEIRA and SERS spectra of ADE in v\_N7 configuration on Au nanostructures composed of 49049 (radius = 6.57 nm) and 104223 (radius = 8.49 nm) (see Fig. S7 in SI). By increasing the NP radius, the AEF and MEF increase by approximately a factor of 2 in SEIRA and 4.5 in SERS; however, all SEIRA and SERS spectra maintain the same profile as that computed for the smallest nanoparticle, demonstrating the robustness of our approach.

Fig. \ref{fig:seira_sers_ade} demonstrates that SEIRA and SERS selectively amplify different vibrational bands, highlighting the complementarity of the two techniques, which has been widely exploited to understand the orientation of adsorbed adenine on metal substrates.\cite{harroun2018controversial} As commented above, three main adsorption models, which depend on the specific binding site of ADE on the Au surface, have been proposed in the literature and validated through various spectroscopic techniques \cite{kundu2008surface,martin2017electrochemical,acres2018experimental,yoshimoto2022three,rodes2009adenine,rueda2012situ,alvarez2014situ,prieto2016quantitative}: N3/N9, N7/N10, and N1/N10.  

\begin{figure}[H]
    \centering
    \includegraphics[width=.8\textwidth]{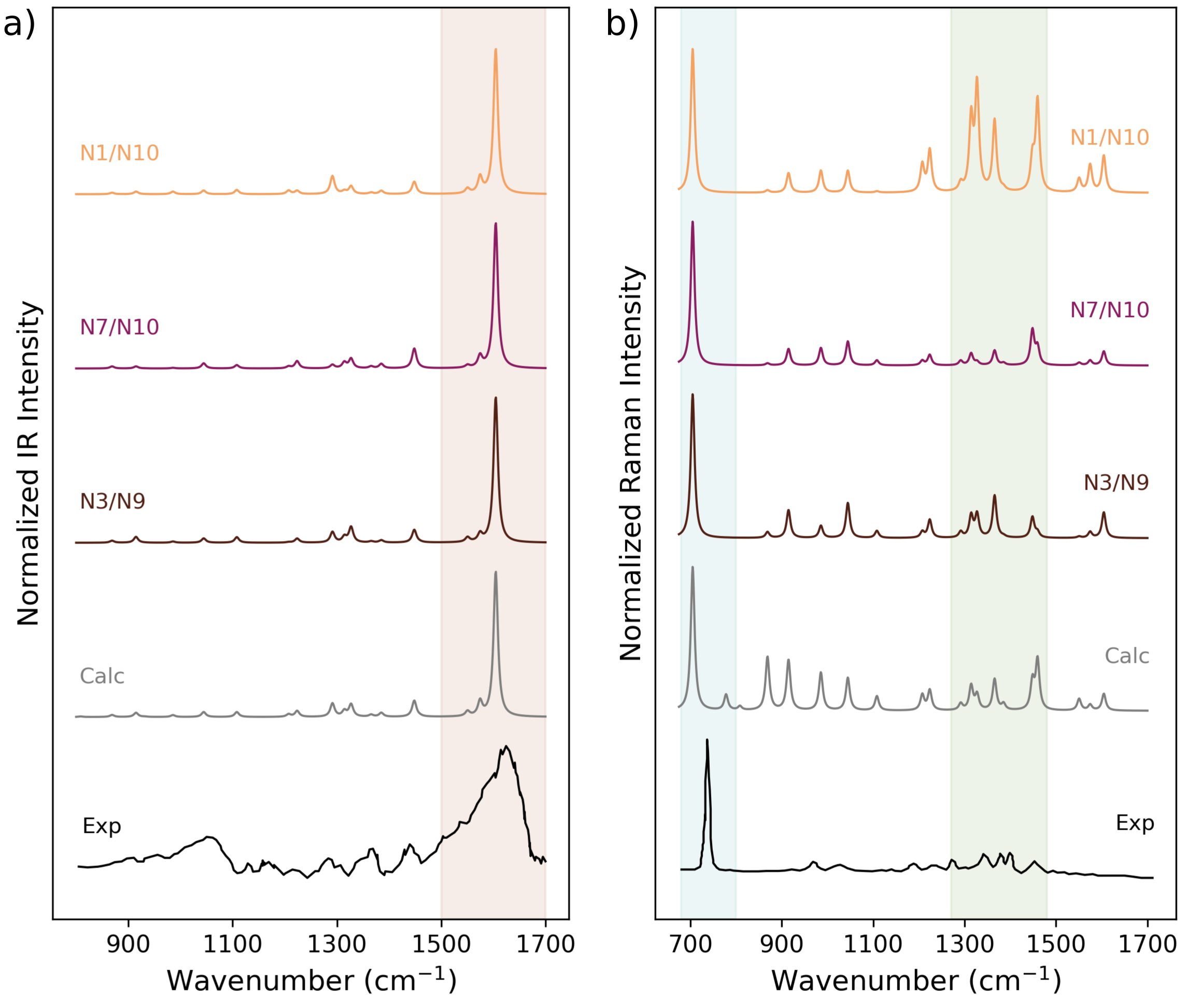}
    \caption{Normalized SEIRA (left panel) and SERS (right panel) spectra, calculated by averaging N3/N9, N7/N10, N1/N10, and all (Calc) configurations of ADE on Au$_{10179}$ Ih, together with the experimental results.\cite{kundu2009adenine}. The characteristic regions of the SEIRA and SERS spectra are highlighted.}
    \label{fig:models_exp}
\end{figure}

To shed light on the most favourable configuration, in Fig. \ref{fig:models_exp}, we average the computed SEIRA and SERS spectra obtained from the corresponding sets of adenine-gold configurations (N1/N10; N7/N10; N3/N9), together with the total averaged computed spectra (Calc), and the experimental data (Exp) of adenine adsorbed on gold nanoshells at neutral pH reported in Ref. \citenum{kundu2009adenine}. 
The experimental SEIRA spectrum (Fig. \ref{fig:models_exp}a, bottom) is dominated by the in-plane symmetric NH$_2$ scissoring mode (1625 cm$^{-1}$), and the in-plane ring modes (1058, 1286, 1310, 1595 cm$^{-1}$), which is generally explained by suggesting that the C6-NH$_2$ group is oriented perpendicularly to the surface with the amino group far from the surface.\cite{kundu2009adenine} The experimental SERS spectrum (Fig. \ref{fig:models_exp}b, bottom) shows a prominent ``ring breathing'' mode at 735 cm$^{-1}$, but it is also characterized by the presence of a spectral fingerprint in the region 1300-1500 cm$^{-1}$ (in particular peaks 1307 and 1337 cm$^{-1}$), and the presence of the weak out-of-plane ring mode at 787 cm$^{-1}$ suggests a slightly tilted adenine orientation.\cite{kundu2009adenine} 

The spectra resulting from the three adsorption models show a slight variation in the relative intensity of the peaks, in particular around 1300 cm$^{-1}$ for SEIRA and within the 1200-1500 cm$^{-1}$ range for SERS, which are associated with the differences in the enhancement profiles discussed above. Based on our results, the N1/N10 configuration appears to be the least accurate in reproducing the experimental findings, as it fails to reproduce the experimental features. By contrast, both the N7/N10 and N3/N9 models provide a satisfactory match with experiment: in particular, the N3/N9 configuration offers the best agreement with the experimental SERS spectrum, successfully reproducing both the ``ring breathing'' band at 735 cm$^{-1}$ and the structured features in the 1300–1500 cm$^{-1}$ region, and its SEIRA spectrum also gives a good agreement with the experimental profile in terms of relative intensity. Our analysis thus suggests that N3/N9 or N7/N10 are the most likely adsorption models. 
Finally, in order to also consider all the configurations that can possibly contribute to the experimental spectrum, we compare the total average spectra, denoted as `Calc', with the experiment (Fig. \ref{fig:models_exp}). This remarkably captures most experimental features, with a good reproduction of the overall spectrum, although some discrepancies are reported, such as the overestimated intensity in the 900–1100 cm$^{-1}$ region of the calculated SERS spectrum.

\subsection{SEIRA spectra of adenine on graphene}

Recently, IR spectra of small aromatic molecules, such as rhodamines and purines, have been recorded on pristine graphene \cite{hu2018versatile}, graphene oxide \cite{hu2019selective}, and carbon dots \cite{hu2019surface}, reflecting the growing interest in graphene and its derivatives as potential SEIRA substrates.\cite{yang2018nanomaterial,hu2019graphene, georgiou2024orientation} This interest arises from the possibility of tuning graphene plasmons to fall in the mid-IR region, i.e. in the region of molecular vibrations, consequently enhancing the corresponding absorption signals.\cite{rodrigo2015mid} To achieve resonance enhancement in the infrared range, the PRF of graphene substrates can be tuned by either changing their size or geometry or adjusting their carrier density through electrical gating or chemical doping, i.e., by varying the Fermi energy $E_{F}$.\cite{hu2019graphene} 

\begin{figure}[!htbp]
    \includegraphics[width=0.55\textwidth]{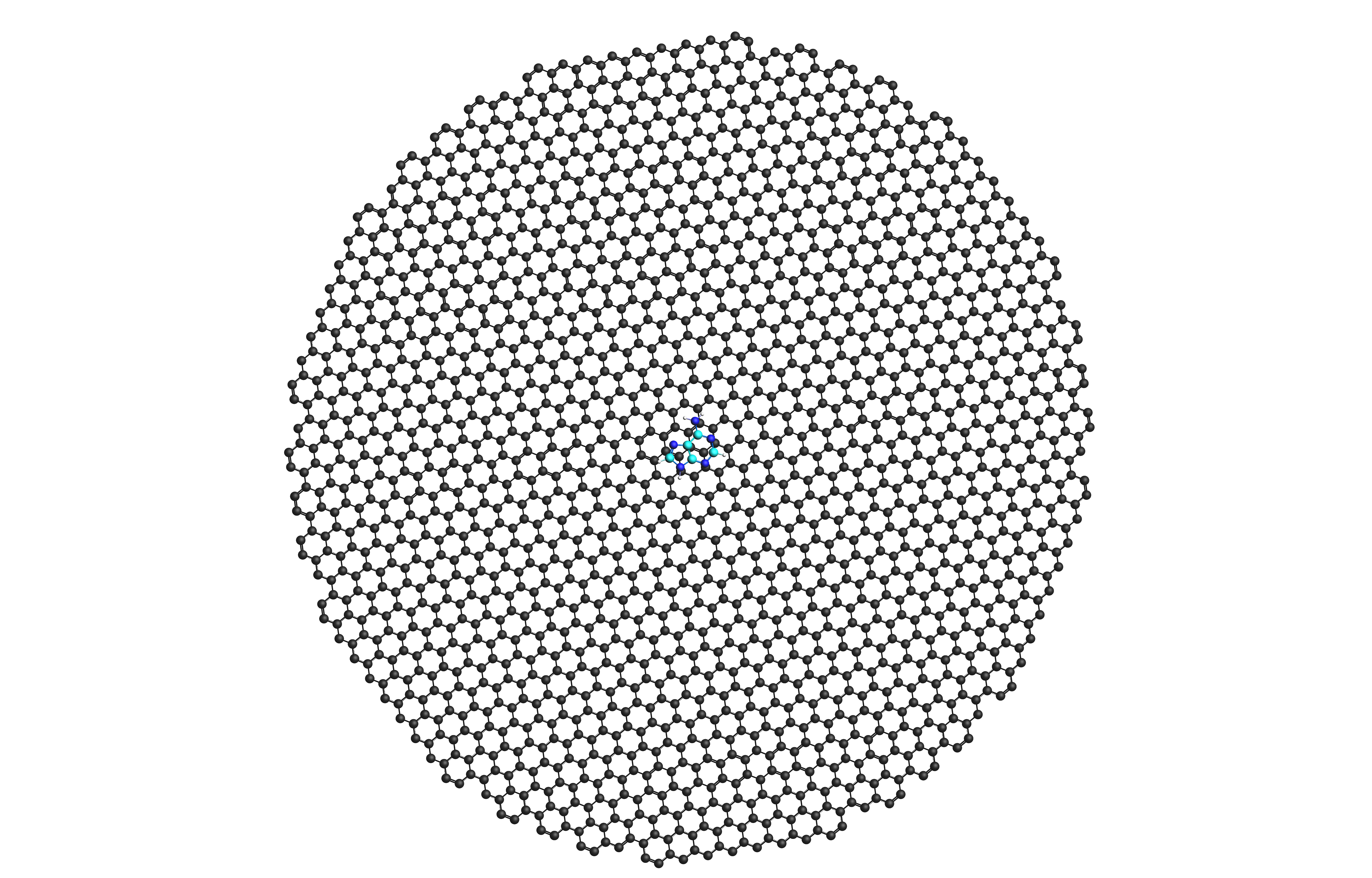} 
    \caption{Graphical depiction of ADE adsorbed on a graphene disk.}
    \label{fig:ade_graphene}
\end{figure}

\begin{figure}[H]
    \centering
    \includegraphics[width=0.50\textwidth]{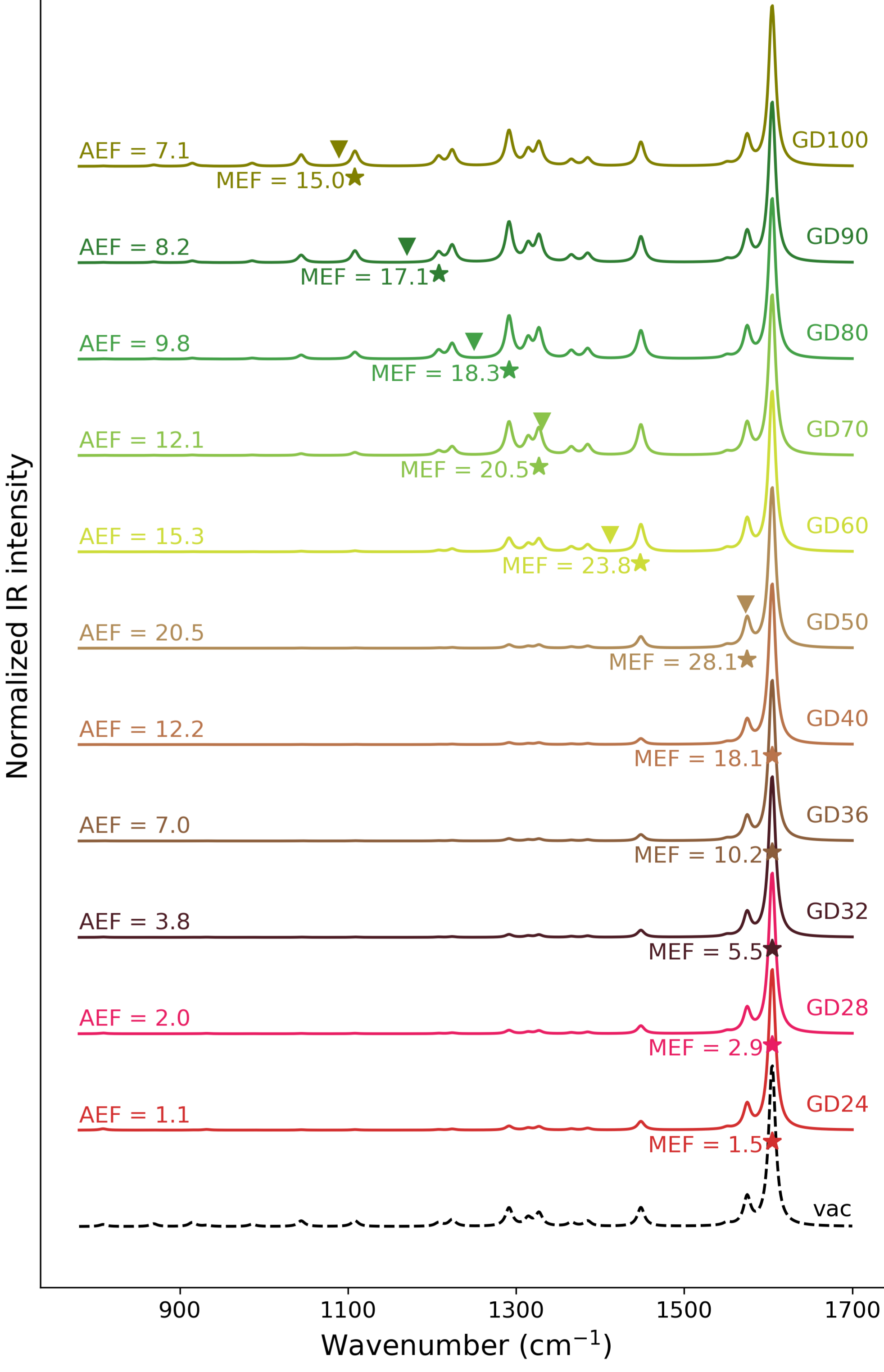}
    \caption{Normalized SEIRA spectra of ADE adsorbed on GDs of increasing size. Stars denote the $i$-MEFs; triangles indicate the PRFs. AEF and MEF values are also reported.} 
    \label{fig:geira_size}
\end{figure}

To showcase the capability of QM/$\omega$FQ to calculate SEIRA spectra, we adsorb ADE on disk-shaped graphene nanostructures with diameters ranging from 24 to 100 nm (GD24 -- GD100). A single adsorption geometry is considered (see Fig. \ref{fig:ade_graphene}), where ADE is placed parallel to the graphene disk, at a distance of 3.5 \AA{}, following experimental and computational evidence.\cite{antony2008structures,voloshina2011theoretical, berland2011van, huang2015molecular} In Fig. \ref{fig:geira_size}, we first focus on the dependence of SEIRA signals on the structural parameters of the graphene substrate (see Table S2 in the SI for the PRF of the various structures). The Fermi energy of the graphene disks is set to 0.4 eV, in line with previous studies.\cite{giovannini2020graphene,bonatti2022silico,lafiosca2023qm}

\begin{figure}[H]
    \centering
    \includegraphics[width=0.98\textwidth]{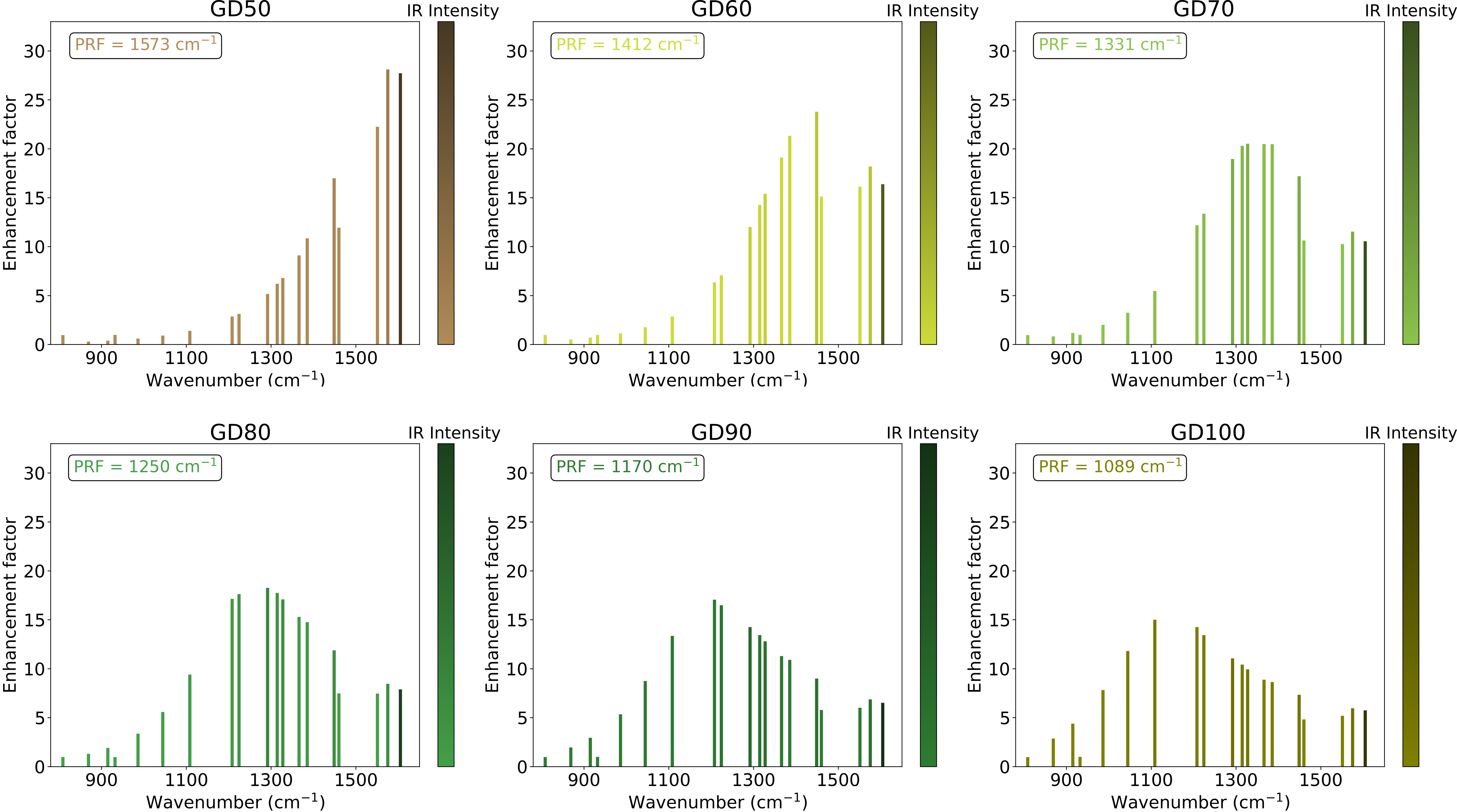}
    \caption{SEIRA enhancement factors computed for each normal mode of ADE on GDs of increasing size. EFs are plotted with a palette following SEIRA intensities.}
    \label{fig:uniform_size}
\end{figure}

The most intense peak for all GD diameters appears at 1605 cm$^{-1}$, which corresponds to the NH$_2$ scissoring and six-membered ring stretching mode, similarly to the gas-phase spectrum (black dashed line). The SEIRA spectra show a clear dependence on GD size. For the smallest GDs (diameter < 40 nm), the IR peaks below 1400 cm$^{-1}$ are characterized by a very low intensity with respect to the gas-phase spectrum, indicating a partial deactivation of ADE's normal modes at lower wavenumbers. As the disk size increases (for diameter > 60 nm), these peaks gradually become more intense, and the spectra of larger disks become increasingly similar to the gas-phase IR spectrum, with its main spectral features largely recovered, even if with different relative intensities. This trend can be rationalized by considering that the most enhanced peaks shift to lower energy as the disk diameter increases, directly following the trend observed for the PRF (see Tables S2 and S4 in SI). In fact, for structures with a diameter < 40 nm, which report the PRF in the range 1734-2258 cm$^{-1}$, the most intense peak at 1605 cm$^{-1}$ also coincides with the $i$-MEF. From GD50 onwards, the $i$-MEF shifts to normal modes that resonate with the PRF of the graphene disk, which progressively redshifts with increasing disk size.\cite{giovannini2020graphene} Such a distribution can better be appreciated by plotting the enhancement factors for each normal mode as a function of the disk size (from GD50 to GD100), which is reported in Fig.\ref{fig:uniform_size}. For larger disks, where the PRF aligns with the considered IR spectral range of adenine, the most enhanced normal modes shift to remain in resonance with the PRF of the respective disk. For example, in GD60, the PRF is 1412 cm$^{-1}$, and the dominant enhancements occur near 1400 cm$^{-1}$, with the $i$-MEF corresponding to the peak at 1448 cm$^{-1}$.

Fig. \ref{fig:geira_size} also reports the average and maximum enhancement factors computed for each structure (see also Fig. S7 and Tab. S4 in the SI). For all the structures, we report an enhancement of the SEIRA signals (AEF > 1.0). 
In particular, the largest AEF (20.5) and MEF (28.1) are reported for GD50. For larger disks than GD50, both AEF and MEF values decrease monotonically (see also Fig. S9 in the SI). These trends can be rationalized by considering that, as the disk radius increases, the induced plasmon field experienced by ADE at the center of the graphene disk becomes more homogeneous. At the same time, ADE moves farther from the disk edges, where the electric field is maximally enhanced. Such effects counterbalance the resonance conditions that are created by enlarging the size of the disk (shifting the PRF to resonate with ADE vibrational modes). This is indeed in line with what has been previously commented on for SERS on graphene disks.\cite{lafiosca2023qm}  

As a final comparison, we note that available experimental data from Ref.\citenum{hu2019selective} report that the IR spectrum of adenine on graphene oxide exhibits an AEF of approximately 30. In particular, the MEF (36.9) is attributed to C=N stretching at 1418 cm$^{-1}$, while other enhanced peaks correspond to C-H bending modes (1241 cm$^{-1}$ and 1394 cm$^{-1}$), C-N stretching (1080 cm$^{-1}$) and NH$_2$ scissoring and aromatic rings stretching mode (1604 cm$^{-1}$).\cite{hu2019selective} Although we focus on an ideal graphene substrate, our results are indeed in good agreement with the experimental values.

\begin{figure}[H]
    \centering
    \includegraphics[width=0.50\textwidth]{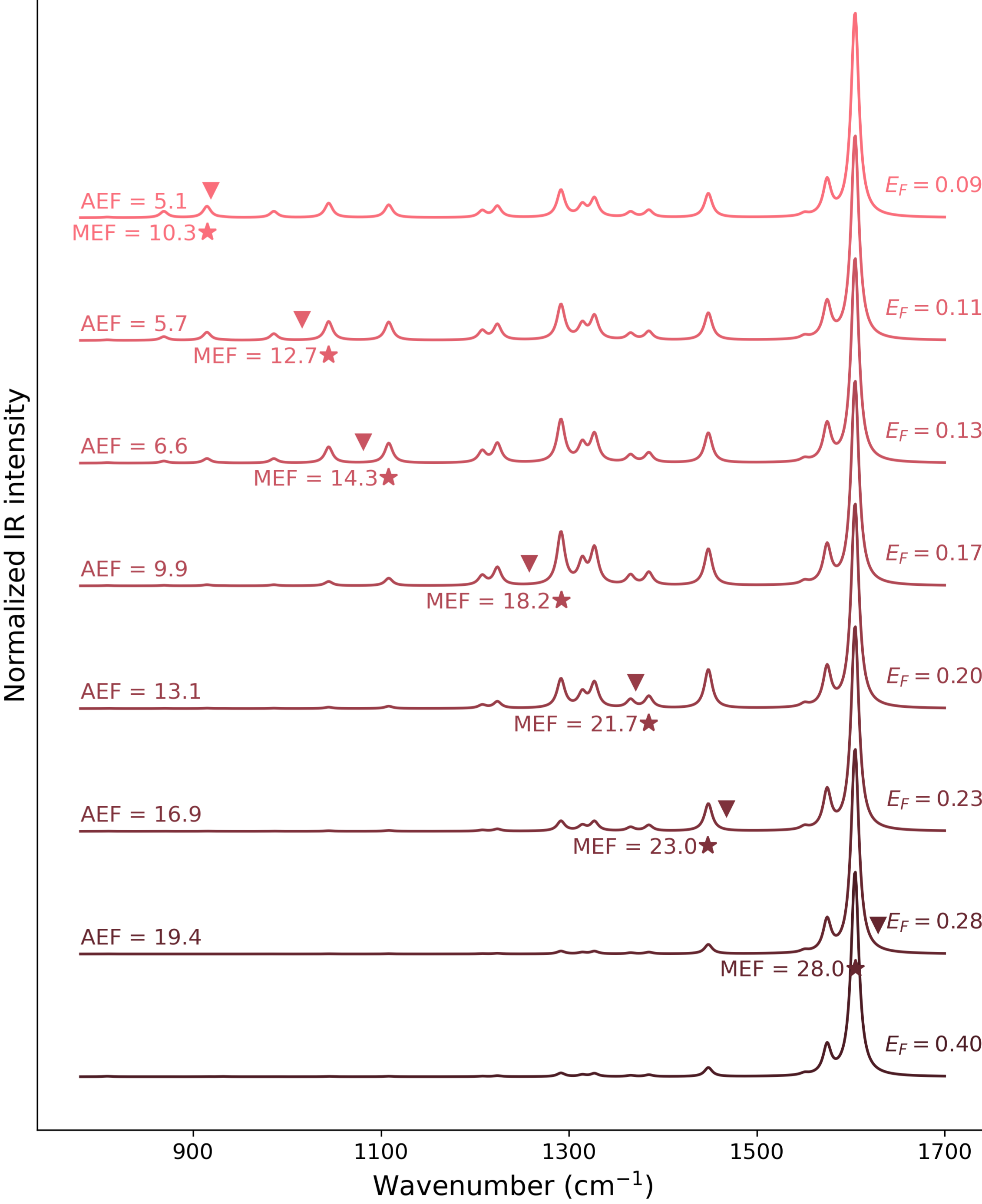}
    \caption{Normalized SEIRA spectra of adenine on GD32 as a function of the Fermi energy (in eV). Stars denote the $i$-MEFs; triangles indicate the PRFs. AEF and MEF values are also reported.}
    \label{fig:gating}
\end{figure}

To conclude this section, we study the SEIRA dependence of ADE adsorbed on a graphene disk with fixed diameter (32 nm) as a function of $E_F$. In particular, we exploit a peculiar property of graphene substrate: by reducing the electron density (i.e., by lowering the Fermi energy), the PRF red-shifts.\cite{giovannini2020graphene, zanotto2023strain}  This allows us to mimic the electronic and optical properties of large nanodisks by using smaller nanostructures, since such a feature is equivalent to increasing the size of the graphene disk. Specifically, we tune the Fermi energy from 0.09 eV to 0.40 eV, which results in the PRF graphically displayed as colored down triangles in Fig. \ref{fig:gating} (see also Table S3 in the SI). These are obtained at the $\omega$FQ level using the open-source plasmonX software.\cite{giovannini2025plasmonx}

As the Fermi energy decreases, the PRF of GD32 progressively redshifts, selectively activating ADE normal modes that resonate with the PRF. The $i$-MEF follows such redshift, while both AEF and MEF gradually decrease (as also highlighted in Tab. S5 in the SI). The maximum AEF (19.4) and MEF (27.9) are observed at $E_{F}$ = 0.28 eV, which are more than five times larger than those obtained when $E_F$ = 0.4 eV (AEF=3.8 and MEF=5.5). This is also reflected by the fact that the spectral shape gradually changes by lowering $E_{F}$, i.e., as the PRF moves toward the IR spectral range of ADE’s vibrational modes. This can also be appreciated by plotting the enhancement factors for all normal modes as a function of $E_F$ (Fig. S10 in the SI). In fact, specific vibrational modes become preferentially enhanced according to their proximity to the PRF. For example, at $E_{F}$ = 0.23–0.20 eV, the PRFs (1468–1371 cm$^{-1}$) enhance modes in the 1300–1500 cm$^{-1}$ region, whereas at $E_{F}$ = 0.13–0.09 eV (PRF 1081–919 cm$^{-1}$), the enhancement shifts toward lower-wavenumber modes with peaks around 900 cm$^{-1}$ emerging distinctly. This is in perfect agreement with the behavior observed for graphene disks of increasing size, highlighting again the direct correlation between PRF and adenine IR modes. 

Interestingly, GD32 with $E_F$ = 0.13 eV and GD100 with $E_F$ = 0.4 eV exhibit nearly identical PRFs (1081 cm$^{-1}$ and 1089 cm$^{-1}$, respectively) with very similar AEF and MEF values (6.58 vs 7.12 and 14.30 vs 15.02), and the same $i$-MEF (at 1108 cm$^{-1}$). This highlights that, by appropriately tuning the Fermi energy of a smaller disk, QM/$\omega$FQ method can effectively replicate the SEIRA spectrum of a molecule adsorbed on a larger graphene disk (GD100, containing roughly ten times more atoms), by using smaller structures (GD32).

\section{Conclusions}

We have presented a fully atomistic multiscale method to simulate the SEIRA spectrum of molecular systems adsorbed on plasmonic substrates. The method couples a QM description of the target molecule with a classical, fully atomistic model for the plasmonic substrate. Specifically, the atomistic electromagnetic models $\omega$FQ and $\omega$FQF$\mu$ describe the plasmonic behavior of noble metal nanostructures and graphene-based materials. 

As a first application, we have computed the SEIRA spectra of the adenine nucleobase adsorbed on gold nanostructures, systematically analyzing the influence of molecular orientation and adsorption site on spectral features and enhancement factors. 
By examining 12 configurations - six on the vertex and six on the face of a icosahedral gold nanoparticle - we have directly compared SEIRA and SERS spectra and enhancements, and benchmarked our results against available experimental spectra. Our analysis reveals that SERS is generally more sensitive than SEIRA, exhibiting higher enhancement factors. Nonetheless, SEIRA and SERS offer complementary spectroscopic information: while SERS is strongly influenced by the adsorption site, which dictates the spectral profile and selectively activates vibrational modes compared to the gas phase, SEIRA is particularly sensitive to the molecular binding configuration. Overall, both the adsorption site and the molecular configuration play a decisive role in shaping the spectral features and modulating enhancement in the two spectroscopies. Following the recent trend on graphene-based materials as SEIRA platforms\cite{oh2021nanophotonic}, we have also explored the SEIRA spectra of ADE on graphene-based nanostructures, by systematically studying the dependence of ADE SEIRA response on the size and Fermi energy of graphene nanodisks, providing insights toward optimizing molecular detection on graphene-based SEIRA platforms.

At this stage, we can already provide a rationalization of experimental results; however, some discrepancies are still present. These can arise from factors that are not currently captured by our current model, e.g., solvent effects and dynamical configurational phase-space sampling. This emphasizes the need for more realistic simulations, including explicit solvent molecules, pH effects, non-ideal substrates, chemical effects,\cite{hu2018versatile,hu2019selective} and a statistically meaningful set of configurations, potentially extracted from molecular dynamics trajectories.\cite{sodomaco2023computational} Addressing these aspects could therefore advance the predictive power of our computational strategy.

\begin{acknowledgement}
This work has received funding from the European Research Council (ERC) under the European Union’s Horizon 2020 research and innovation programme (grant agreement No. 818064). We gratefully acknowledge the Center for High Performance Computing (CHPC) at SNS for providing the computational infrastructure. 
\end{acknowledgement}

\begin{suppinfo}
SEIRA dependence on DFT grid quality and differentiation step. Effect of QM/FQF$\mu$ contribution into the Hamiltonian. ADE normal modes. Raw data of enhancement factors for each ADE-substrate. SEIRA dependence on the size of Au NP. PRF values of GD as a function of the radius and Fermi energy. 
\end{suppinfo}

\bibliography{biblio.bib}
\pagebreak

\end{document}